\documentclass[12pt]{article}
\usepackage{amsmath,latexsym}
\usepackage{graphicx}
\begin{document}

 \title{\Huge Discussions on a special static spherically symmetric perfect fluid
 solution of Einstein's equations
   }
 \author{F.Rahaman$^*$, M.Kalam$^{\dag}$, S. Chakraborty$^*$ , K.Maity$^*$ and B. Raychaudhuri$^{\ddag}$  }

\date{}
 \maketitle

 $ $

 $ $

 \begin{abstract}
In this article, a special  static spherically
symmetric perfect fluid solution of Einstein's equations is provided. Though
pressure and density both diverge at the origin,  their ratio
remains constant. The  solution presented here fails to give positive pressure but
nevertheless, it satisfies all energy conditions. In this new
spacetime geometry, the metric becomes singular  at some finite value of
radial coordinate although, by using isotropic coordinates, this singularity could be avoided, as has been shown here. Some characteristics of this solution are also discussed.

$ $

$ $ \textbf{Key words:}  Einstein's gravitational field equations; Exact solution; energy conditions.

\end{abstract}
$ $

 $ $

 $ $

 $ $

%\bigskip
 %\medskip
  \footnotetext{
 $*$Dept.of Mathematics, Jadavpur University, Kolkata-700 032, India

                                  E-Mail:farook\_rahaman@yahoo.com\\
$\dag$Dept. of Phys. , Netaji Nagar College for Women, Regent Estate, Kolkata-700092, India.\\
E-Mail:mehedikalam@yahoo.co.in\\
$\ddag$Dept. of Phys. , Surya Sen Mahavidyalaya, Siliguri, West Bengal, India :\\
 E-mail: biplab.raychaudhuri@gmail.com

}
    \mbox{} \hspace{.2in}

\section{Introduction}
Einstein's gravitational field equations are 
non linear in nature. Physicists have been trying to obtain exact
analytical form of the interior perfect fluid solutions for
various reasons such as modelling of stars, describing
astrophysical phenomena etc. Delgaty and Lake [1] have
provided an excellent review of static spherically symmetric
perfect fluid solutions of Einstein's equations. Among the
solutions existing in literature,  Schwarzschild
interior solution with uniform density is the most studied for its simplicity. Nevertheless, exact
solutions always demand importance for  understanding of the
inherent non-linear distinguishing peculiarity of gravity. In this
article, we provide a discussion on a special  class of exact
static spherically symmetric perfect fluid solutions of Einstein's
equations.

\section{The basic equations and their integrals}
The static spherically symmetric metric in Schwarzschild coordinates $ ( t,r,\theta, \phi ) $ can be written as 

\begin{equation}
               ds^2=  - e^{\nu(r)} dt^2+ e^{\lambda(r)} dr^2+r^2( d\theta^2+\sin^2\theta
               d\phi^2)
          \end{equation}

Field equations pertinent to this metric are

\begin{equation}
e^{-\lambda} \left[\frac{\lambda^\prime}{r} - \frac{1}{r^2}\right]+\frac{1}{r^2}= 8\pi \rho,
\end{equation}

\begin{equation}e^{-\lambda}
\left [\frac{1}{r^2}+\frac{\nu^\prime}{r}\right]-\frac{1}{r^2}=8\pi p,
\end{equation}

\begin{equation}\frac{1}{2} e^{-\lambda}
\left[\frac{1}{2}(\nu^\prime)^2+ \nu^{\prime\prime}
-\frac{1}{2}\lambda^\prime\nu^\prime + \frac{1}{r}({\nu^\prime-
\lambda^\prime})\right] =8\pi p.
\end{equation}
We use prime to represent derivative with respect to the radial coordinate.

The usual conservation equation $ T^{ik}_{;k} = 0 $ implies
\begin{equation}
               p^\prime = - ( p + \rho) \frac{\nu^\prime}{2}
                   \end{equation}

Note that in the above set of four equations there are  four unknowns, so in
principle, there should exist unique exact solution. They are found to be
\begin{equation}
               e^\nu  =\frac{K}{ r}
               \end{equation}
\begin{equation}
               e^\lambda  = \frac {1} {\left(\frac{C}{r} -
               4\right)}
          \end{equation}
\begin{equation}
              p  = -  \frac{1}{8 \pi r^2}
               \end{equation}
          \begin{equation}
              \rho  =   \frac{5}{8 \pi r^2}
                \end{equation}
where $K$ and $C$ are arbitrary constants.

Two plots are given below representing the dependence of metric components as functions of radial coordinate with $K$ and $C$ taken as parameters. Pressure and density are also plotted to show the nature of their radial dependence.
\begin{figure}[htbp]
    \centering
        \includegraphics[scale=.5]{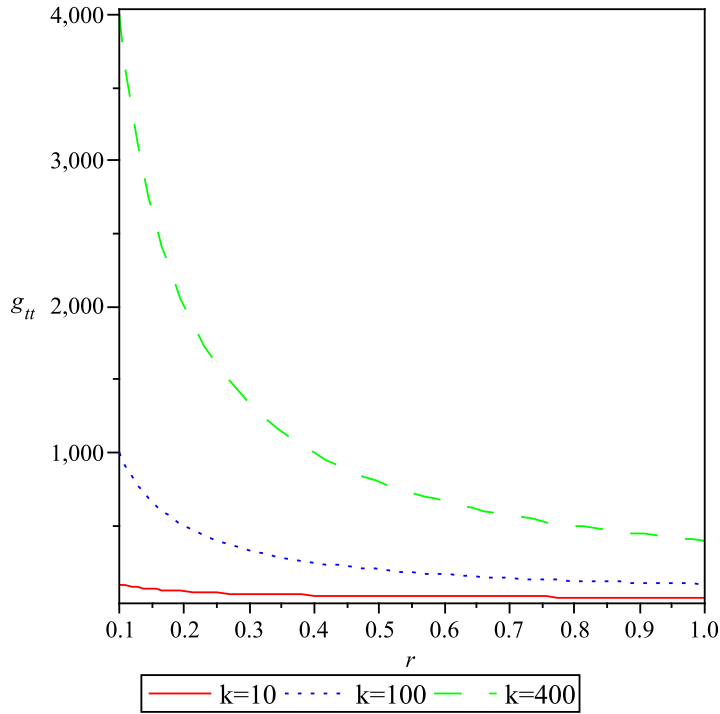}
        \caption{$g_{tt}$ vs $r$ for different values of $K$. }
  \end{figure}
\begin{figure}[htbp]
    \centering
        \includegraphics[scale=.5]{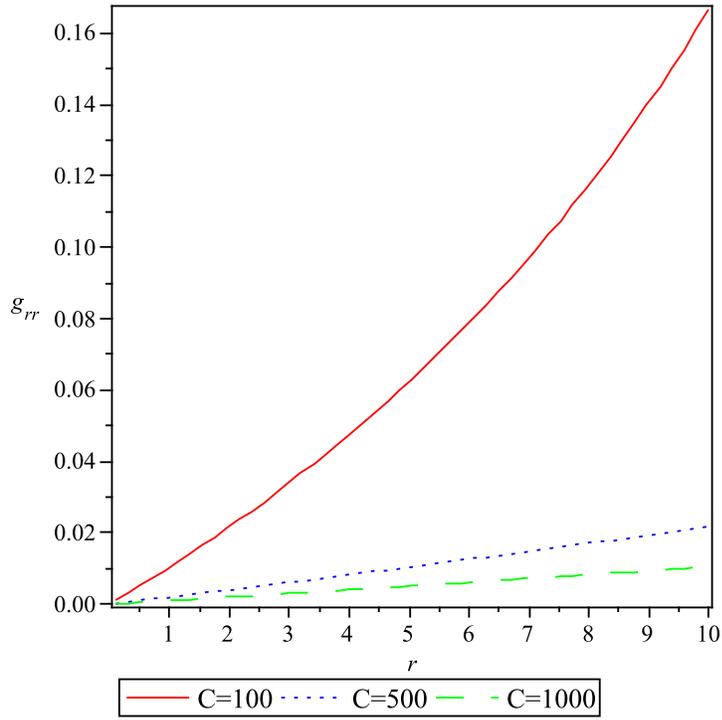}
        \caption{$g_{rr}$ vs $r$ for different values of $C$. }
 \end{figure}

 The pressure and density both diverge at the origin but their
ratio remains constant i.e. $\frac{p}{\rho} = - \frac{1}{5}$. Here
$\mid p \mid$ and $\rho$ are monotonic decreasing functions of r.
One can also see that  $ \rho > 0$, $p+ \rho > 0$, $  \rho + 3p
> 0$ as well as $ \rho > \mid p \mid $. Thus all energy conditions
including dominant energy condition are satisfied.

\begin{figure}[htbp]
    \centering
        \includegraphics[scale=.5]{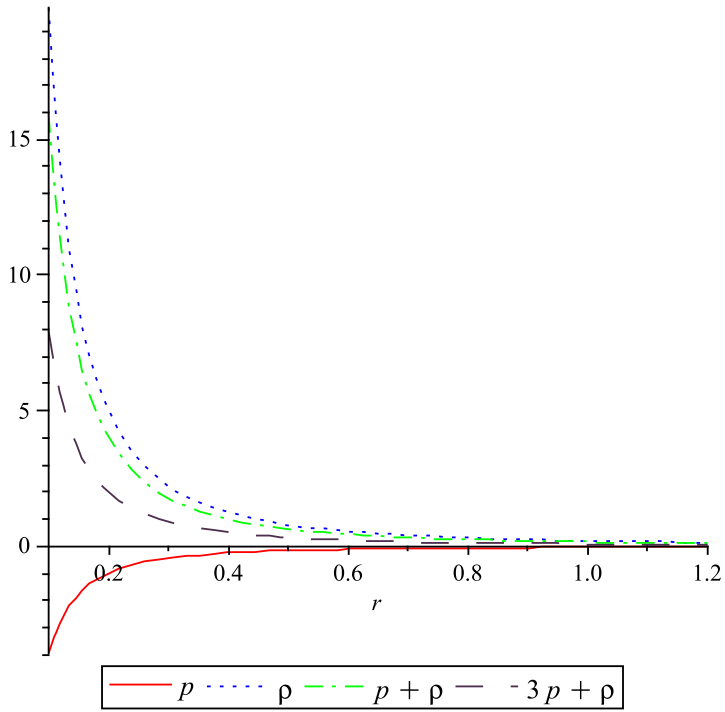}
        \caption{$p$ and $\rho$ vs $r$}
\end{figure}

 One of the important features of Einstein's equations is the appearance of curvature singularity. In this solution there is a coordinate singularity at $r=C/4$. But a close look at this spacetime solution shows that the metric does not contain any curvature singularity apart from the well known $r=0$ singularity. (Later we shall show that in suitably chosen isotropic coordinates this singularity does not appear.) This can  easily be seen by calculating the analytical expressions of the curvature invariants which, for the present metric are given by
\begin{equation} 
R = \frac{8}{r^2}. 
\end{equation}
\begin{equation} 
R_{\mu\nu}R^{\mu\nu} = \frac{28}{r^4}.
 \end{equation}
\begin{equation} 
R_{\mu\nu\alpha\beta}R^{\mu\nu\alpha\beta} = \frac{168}{r^4}
- \frac{80C}{r^5} + \frac{12C^2}{r^6}. 
\end{equation}
 The expressions show that they are regular everywhere and diverge only at $r=0$. We
also see that the Kretschmann scalar
$R_{\mu\nu\alpha\beta}R^{\mu\nu\alpha\beta}$ becomes zero as $r \rightarrow \infty $. Hence the solution is asymptotically well behaved. The dependence of the scalars on the radial variable is shown in representative curve below. We avoid the  $C/4$ coordinate singularity by suitably choosing the range of the radial coordiate. 
\begin{figure}[htbp]
    \centering
      \includegraphics[scale=.5]{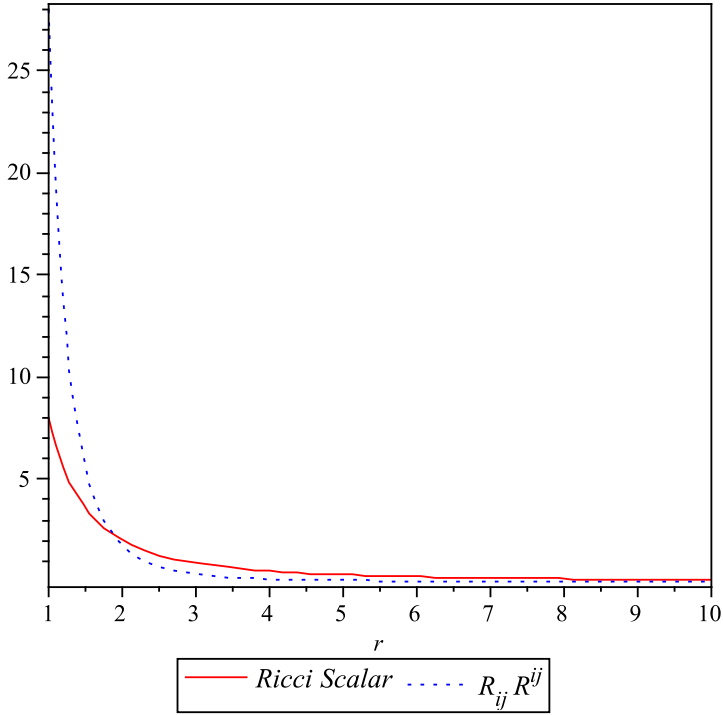}
       \caption{$R$ and $R_{\mu\nu}R^{\mu\nu}$ vs $r$. }
     \end{figure}

\begin{figure}[htbp]
    \centering
        \includegraphics[scale=.5]{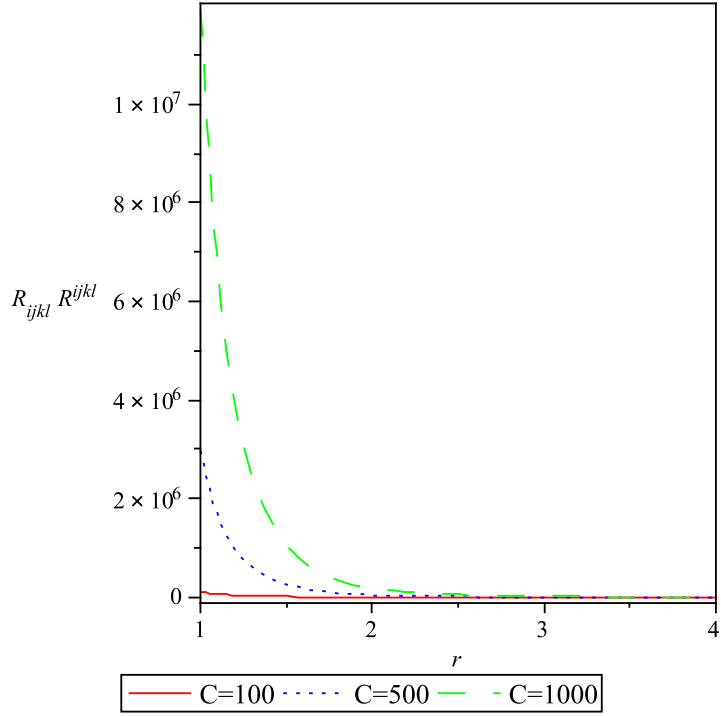}
        \caption{Kretschmann scalar $R_{\mu\nu\alpha\beta}R^{\mu\nu\alpha\beta}$ vs $r$ for different values of $C$. }
   \end{figure}

\section{Isotropic coordinates}
The new interior metric is now written in isotropic coordinate as given by

\begin{equation}
               ds^2=  - \left(\frac{K}{r}\right) dt^2+ [\chi (\sigma)]^2 \left[
d\sigma^2+\sigma^2( d\theta^2+\sin^2\theta d\phi^2)\right].
\end{equation}

A straighforward comparison with the static spherically symmetric metric (Eq.~(1)) yields

\begin{equation}
               r^2= \chi^2 \sigma^2,
\end{equation}
 and
          \begin{equation}
             \frac{dr^2}{ \left( \frac{C}{r} - 4\right)} = \chi^2 d\sigma^2.
                   \end{equation}

The above two equations yield
\begin{equation}
               r = \frac{1}{8} \left[ C + C \sin ( \ln
               \frac{\sigma}{\sigma_0}) \right],
              \end{equation}

\begin{equation}
               \chi^2  = \frac{ \left[ C + C \sin ( \ln
               \frac{\sigma}{\sigma_0}) \right]^2}{64\sigma^2}.
                \end{equation}
         where $\sigma_0$ is an integration constant.

Hence, finally, the metric takes the form as
\begin{equation}
               ds^2=  - \frac{8K  dt^2}{\left[ C + C \sin ( \ln
               \frac{\sigma}{\sigma_0}) \right]} + \frac{ \left[ C + C \sin ( \ln
               \frac{\sigma}{\sigma_0}) \right]^2}{64\sigma^2} \left[ 
d\sigma^2+\sigma^2( d\theta^2+\sin^2\theta d\phi^2)\right]
                   \end{equation}

 Note that in  isotropic coordinate system, the singularity $r = \frac{C}{4}$ does not appear. This supports our earlier assertion that this singularity is a coordinate one which was evident from the calculations of the invariants.

\section{Matching with exterior Schwarzschild solution}
Now, we match our interior solution to the exterior Schwarzschild
solution at a junction interface $(S)$ situated at $r=a$. We
impose the continuity of $g_{\mu\nu}$ across the surface S:

 $g_{\mu\nu (int)}\mid _S$ = $g_{\mu\nu(ext)}\mid _S$

 at $ r=a$ [ i.e. on the surface S ].

 The continuity of the metric then gives generally

$ {e^{\nu}}_{int}(a) = {e^{\nu}}_{ext}(a) $ and $
{e^{\lambda}}_{int}(a) = {e^{\lambda}}_{ext}(a) $.

Hence one can find

\begin{equation}
\frac{K}{a}= \left( 1 - \frac{2M}{a}\right) 
\end{equation}

and
\begin{equation} 
\left( \frac{C}{a} - 4 \right) = \left( 1 - \frac{2M}{a}\right)  
\end{equation}
These imply,
\begin{equation} 
K  =  a - 2M 
\end{equation}
  \begin{equation}
C =  5a - 2M   
\end{equation}

  Since  $ r < \frac{C}{4} $, so to match  our interior solution to the exterior Schwarzschild
solution at a junction interface $(S)$ situated at $r=a$, one has
to take  $ a < \frac{C}{4} $. Equations (22) and (23) confirm this
i.e. $ a < \frac{C}{4} $.

 Hence, our interior metric takes the form as

\begin{equation}
ds^2=  - \left( \frac{a - 2M}{r}\right) dt^2+ \frac{ dr^2}{\left( 
\frac{5a - 2M}{r} - 4\right)} +r^2( d\theta^2+\sin^2\theta   d\phi^2)
 \end{equation}

We will face a problem to get the boundary surface ( thin shell $S$
) as  $p|_{ r=a }$  $= 0$, but p never vanishes except at
infinity. We see that the metric coefficients continuous at the
junction i.e. at $S$. However, the metric need not be differentiable
at the junction and the affine connection may be discontinuous
there. This statement may be quantified in terms of second
fundamental form of the boundary.

The second fundamental forms associated with the two
sides of the shell are~\cite{israel,rahaman}
\begin{equation}
K_{ij}^\pm =  - n_\nu^\pm\ \left[ \frac{\partial^2X_\nu}
{\partial \xi^i\partial \xi^j } +
 \Gamma_{\alpha\beta}^\nu \frac{\partial X^\alpha}{\partial \xi^i}
 \frac{\partial X^\beta}{\partial \xi^j }\right]|_S 
\end{equation}
where $ n_\nu^\pm\ $ are the unit normals to $S$,
\begin{equation} n_\nu^\pm =  \pm   | g^{\alpha\beta}\frac{\partial f}{\partial X^\alpha}
 \frac{\partial f}{\partial X^\beta} |^{-\frac{1}{2}} \frac{\partial f}{\partial X^\nu} \end{equation}
with $ n^\mu n_\mu = 1 $.

 $\xi^i$ are the intrinsic coordinates on the shell with $f =0 $ is the parametric equation of the shell S and $-$  and  $ + $ corresponds to interior (our) and exterior (Schwarzschild ). \textit{Since the shell is infinitesimally thin in the radial direction there is no radial pressure}. Using Lanczos equations~\cite{israel,rahaman}, one can find the surface energy term
$\Sigma$ and surface tangential pressures $ p_\theta = p_\phi
\equiv p_t $ as
$$ \Sigma =  - \frac{1}{4\pi a}[ \sqrt{e^{-\lambda}}]_-^+ ,$$
$$     p_t =   \frac{1}{8\pi a}[ ( 1 + \frac{a \nu^\prime }{2})
\sqrt{e^{-\lambda}}]_-^+ .$$

The metric functions are continuous on S, then one finds
\begin{equation}  \Sigma = 0 \end{equation}
 and
\begin{equation}    p_t = \frac{1}{16\pi a} \left[ \frac{2M}{ a
({\sqrt{1 - \frac{2M}{a}})}} + \sqrt{ \frac{C}{a} - 4}\right]
\end{equation}
Hence one can match our interior solution with an exterior
Schwarzschild solution in the presence of a thin shell. The  whole
spacetime is given by our metric  and Schwarzschild metric which
are joined smoothly.

Here the Tolman-Whittaker expression for the active gravitational
mass is given by
\begin{equation} M_G = \int_0^{r} 4 \left( T_0^0 - T_1^1 - T_2^2 - T_3^3 \right )
e^{\frac{(\nu + \lambda )}{2}} r^2 dr = 2\sqrt{K}\left[
\sqrt{C}-\sqrt{C-4r}\right]
\end{equation}
The expression for active gravitational mass in equation(28)
clearly shows that a radial dependance which is increasing
function of r. We observe from the equation (28) that as $r
\rightarrow 0$, $M_G \rightarrow 0$. In the interior region the
maximum active gravitational mass is given by
\begin{equation} M_G (Max)  = 2\sqrt{K}\left[
\sqrt{C}-\sqrt{C-4a}\right]
\end{equation}
As $a\rightarrow \frac{C}{4}$, $M_G (Max)\rightarrow 2\sqrt{KC} $.
Though the pressure and density both diverge at origin but active
gravitational mass tends to zero as $r \rightarrow 0$. Thus the
active gravitational mass does not suffer the well known problem
of singularity.

\begin{figure}[htbp]
    \centering
        \includegraphics[scale=.9]{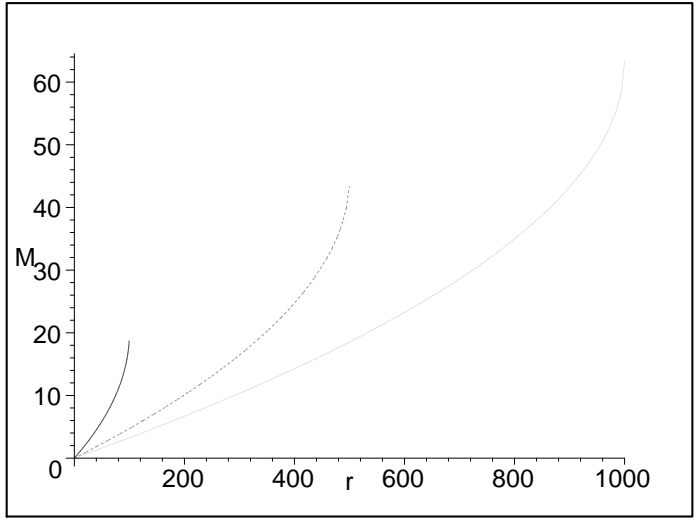}
        \caption{$M_G$ vs r for different values of C( solid line for $C=100$, dotted line for
        $C=500$ and shaded line for $C=1000$. }
    \end{figure}

\section{Concluding Remarks}
We are now in position to summarize our findings. 
\begin{enumerate}
\item One can note that at the singularity $r=\frac{C}{4}$, $g_{tt}$ does not vanish and this
implies that no horizon exists. Also all the curvature invariants are regular everywhere implying that the singularity appearing in (7) is only a coordinate singularity.
\item Our results failed to give positive pressure but  all the
energy conditions are satisfied for the physical acceptability of
perfect fluid source. Also pressure and density failed to be
regular at the origin but their ration remains constant and active
gravitational mass always positive and will vanish as $r
\rightarrow 0$ i.e. it does not have to tolerate the problem of
singularity.
\item  Though our results failed to get boundary surface where
interior solution will match with exterior Schwarzschild solution,
in spite of,  we have shown that there exists a thin shell (boundary surface ) where interior metric and Schwarzschild metric are joined smoothly.
\item One of the elementary criteria for physically acceptability
spherically symmetric solution is that the subluminal sound speed
to be less than unity. Here, we find that the numerical value of
the subluminal sound speed , $ \mid v_s^2 \mid  =  \mid \frac{dp}{d\rho}\mid  <  1 $.
\end{enumerate}

The discussion above refers to the importance of the present solution in the field of
physically acceptable static spherically symmetric perfect fluid solutions of Einstein equations.

\textbf{Note added: } After completing this work, we are informed
that our solution is a special case of the solutions obtained by B
Kuchowicz~\cite{kuchowicz}.

\end{document}